\documentclass{article}

\PassOptionsToPackage{numbers, compress}{natbib}
\usepackage[main, final]{neurips_2025}

\usepackage[utf8]{inputenc} % allow utf-8 input
\usepackage[T1]{fontenc}    % use 8-bit T1 fonts
\usepackage{url}            % simple URL typesetting
\usepackage{booktabs}       % professional-quality tables
\usepackage{amsfonts}       % blackboard math symbols
\usepackage{nicefrac}       % compact symbols for 1/2, etc.
\usepackage{microtype}      % microtypography
\usepackage{xcolor}         % colors
\usepackage{graphicx}

\usepackage{amsmath}
\usepackage{amsfonts}
\usepackage[misc]{ifsym}
\usepackage{hyperref}       % hyperlinks
\title{SemanticVocoder: Bridging Audio Generation and Audio Understanding via Semantic Latents}

\author{%
    Zeyu Xie$^{1,2}$, Chenxing Li$^2$, Qiao Jin$^1$, Xuenan Xu$^3$, Guanrou Yang$^{2,4}$\\
        \textbf{Wenfu Wang$^2$,  Mengyue Wu$^{4}$, Dong Yu$^{2}$, Yuexian Zou$^{1,\textrm{\Letter}}$ }\\
  $^1$Peking University, Shenzhen
  $^2$Tencent AI Lab, Shenzhen\\
  $^3$Shanghai AI Lab, Shanghai
  $^4$SJTU, Shanghai\\
  \textit{zeyuxie25@stu.pku.edu.cn}, \textit{zouyx@pku.edu.cn}
}
\begin{document}
\maketitle
\begin{abstract}
Recent audio generation models typically rely on  Variational Autoencoders (VAEs) and perform generation within the VAE latent space.
Although VAEs excel at compression and reconstruction, their latents inherently encode low-level acoustic details rather than semantically discriminative information, leading to entangled event semantics and complicating the training of generative models.
To address these issues, we discard VAE acoustic latents and introduce semantic encoder latents, thereby proposing SemanticVocoder, a generative vocoder that directly synthesizes waveforms from semantic latents.
Equipped with SemanticVocoder, our text-to-audio generation model achieves a Fréchet Distance of 12.823 and a Fréchet Audio Distance of 1.709 on the AudioCaps test set, as the introduced semantic latents exhibit superior discriminability compared to acoustic VAE latents.
Beyond improved generation performance, it also serves as a promising attempt towards unifying audio understanding and generation within a shared semantic space.
Generated samples are available at~\href{https://zeyuxie29.github.io/SemanticVocoder/}{\textcolor{cyan}{\textit{https://zeyuxie29.github.io/SemanticVocoder/}}}.
% These demonstrate that SemanticVocoder provides a promising trial toward unifying audio understanding and generation within a shared semantic space.
\end{abstract}
\begin{figure}[htbp]
    \vspace{-10pt} \centerline{\includegraphics[width=0.50\linewidth]{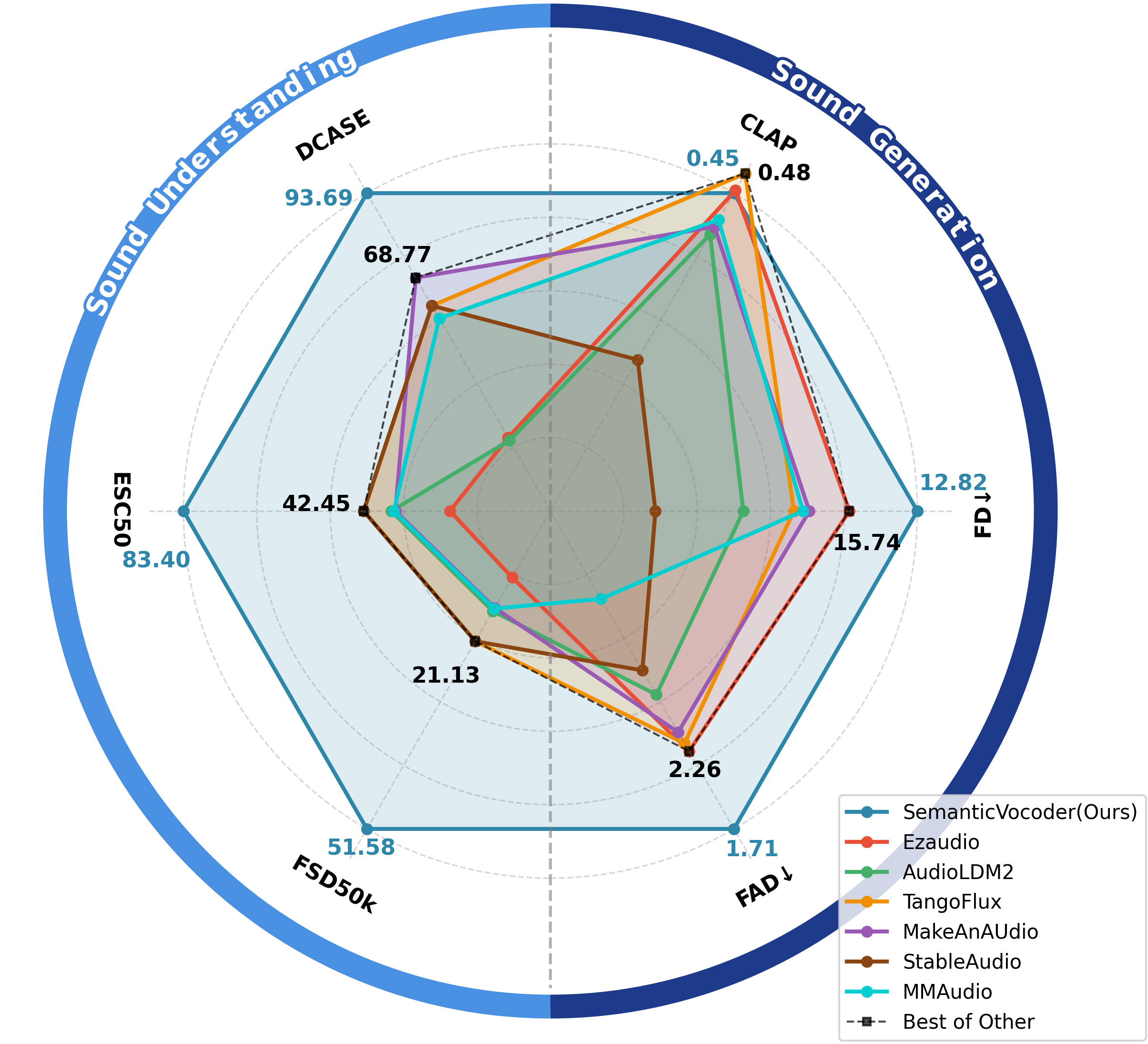}}
    \caption{
    The SemanticVocoder pioneers the generation of waveforms directly from semantic latents, thereby bridging understanding-oriented representations and generation tasks.
    {\color[HTML]{4A90E2}\bf{(Left)}:}
    Three sub-tasks from the HEAR benchmark are employed to evaluate the latent representations, in which linear classifiers are trained on fixed latents.
    % Three sub‑tasks from the HEAR benchmark are employed to evaluate the latent representations by training linear classification layers.
    %The introduced semantic latents 
    The semantic latents exhibit a more discriminative semantic structure than the acoustic VAE latents used in previous work.
    {\color[HTML]{1E3A8A}\bf{(Right)}:} 
    For the downstream text-to-audio task, a text-to-latent model predicts latents conditioned on input text.
    The predicted latents are then fed into SemanticVocoder for audio synthesis, yielding superior performance.
    %Downstream generation tasks demonstrate that employing semantic latents for audio generation with the SemanticVocoder achieves superior performance.
    }
%An illustration of .}
\label{fig:radar}
\end{figure}

\section{Introduction}
\label{sec:intro}
Text-to-audio (TTA) generation has attracted considerable attention and advanced rapidly in recent years.
Previous audio generation systems~\cite{hai2024ezaudio, liu2023audioldm2,  hung2024tangoflux, huang2023make1,  evans2025stable, cheng2025mmaudio} typically follow the classic Latent Diffusion Model (LDM) architecture, which relies on a first-stage Variational Autoencoder (VAE).
The VAE compresses audio into compact latents via an encoder and reconstructs the original audio through a \underline{latent-to-waveform} decoder, while a second-stage \underline{text-to-latent} generative model performs prediction in this latent space.
Because the VAE is fundamentally based on pure acoustic compression, its latent representation is referred to as \textbf{acoustic latents}.

However, acoustic latents pose challenges for second-stage generative models.
The reconstruction objective compels the VAE encoder to retain fine-grained acoustic details within the latent space. 
This preservation ensures high-fidelity reconstruction but yields acoustically dense latents, leading to weak semantic discriminability.
Moreover, the VAE bottleneck typically employs low-dimensional representations, imposing an upper bound on representation capacity and further limiting semantic information capture (see Table \ref{tab:understanding}). 
%As a consequence, The second-stage models must learn to map complex, low-dimensional and low-level acoustic variations rather than focusing on higher-level semantic structures. 
Consequently, second-stage models face a challenging cross-modal task: they must map semantic textual captions directly to complex, low-level acoustic variations—an objective substantially more difficult than mapping to high-level semantic structures.
These properties are suboptimal for generative modeling

We thus attempt to introduce a novel representation to mitigate the adverse effects of VAE acoustic latents for audio generation.
An intuitive ideal representation is that extracted by a semantic encoder, which we refer to as \textbf{semantic latents}. 
% This is because the training objective of the semantic encoder compels the latents to possess strong semantic disentanglement and a clear discriminative structure to achieve favorable performance in downstream understanding tasks. 
To excel in downstream understanding tasks, semantic encoders are optimized to learn latent spaces characterized by strong semantic disentanglement and clear discriminative structure.
% Such latents typically exhibit high-dimensional, abstract, and semantically rich characteristics, which enable them to capture the ``content'' of audio (e.g., ``a dog barking'') rather than being confined to acoustic details, and this structure is more conducive to generative model modeling.
These high-dimensional, abstract, and semantically rich latents effectively capture the high-level ``content'' of the audio (e.g., ``a dog barking'') rather than fine-grained acoustic details, making them inherently more conducive to generative modeling.

However, directly incorporating semantic latents into the classic LDM framework is non-trivial, as such latents prioritize semantic information at the expense of acoustic details, resulting in severe audio distortion when reconstructed  through traditional VAE training frameworks.
We thus propose the \textbf{SemanticVocoder}, a flow-matching approach that directly synthesizes waveforms from semantic latents in a generative manner. 
It exhibits the following properties: 
(1) It leverages high-dimensional semantic latents without dimensionality reduction, thereby preserving well-structured semantic representations.
%This mitigates the adverse effects of acoustic redundancy and the low-dimensional bottleneck inherent to VAE latents.
This mitigates issues arising from acoustic redundancy and the low-dimensional bottleneck inherent in VAE latents.
(2) It shifts the training paradigm from VAE-based reconstruction to flow-matching driven generation, thus alleviating the objective mismatch between VAE reconstruction and second-stage generation in the original framework. 
(3) With semantic latents as the anchor, the text-to-latent model and SemanticVocoder can be trained simultaneously, rendering the two models mutually independent and endowing them with plug-and-play capabilities. 
By contrast, conventional VAEs and second-stage models must be trained sequentially, as the VAE latents vary during VAE training. 
Our contributions are summarized as follows:
\begin{enumerate}

\item We propose SemanticVocoder, which leverages the semantic latents to \textbf{directly generate waveforms}, enabling the audio generation framework to operate in semantic latent space, while \textbf{eliminating any reliance on VAE modules} and alleviating their negative impacts.
\item  By incorporating SemanticVocoder, our text-to-audio system achieves excellent performance on AudioCaps, with a Fréchet Distance of 12.823 and a Fréchet Audio Distance of 1.709.
\item SemanticVocoder bridges semantic latents and generation tasks, thereby enabling semantic latents to support unified modeling for both audio generation and audio understanding.
\end{enumerate}

\section{Related works}

\label{sec:related_word}

% \subsection{Neural vocoder and audio VAE}
\textbf{Neural vocoder and audio VAE}
Neural vocoders, such as HiFi-GAN~\cite{kong2020hifi} and BigVGAN~\cite{lee2022bigvgan}, typically take intermediate acoustic representations such as mel-spectrograms or audio codec tokens as input, and reconstruct waveforms.
On the other hand, VAEs~\cite{evans2025stable, kumar2024high} follow an encoding–decoding paradigm: they first compress original audio into compact latent representations via an encoder, reconstruct waveforms from these latents through a decoder.
Both traditional neural vocoders and VAEs rely exclusively on acoustic representations, which are reconstruction-oriented; in contrast, the proposed SemanticVocoder utilizes semantic latents and is oriented toward generation.
% Neural vocoders, like HiFi-GAN~\cite{kong2020hifi} and BigVGAN~\cite{lee2022bigvgan}, aim to convert spectral features or audio codec tokens into waveforms.
% %(e.g., waveform-based ones, as well as spectrogram-based ones combined with neural vocoders)
% While audio VAEs~\cite{evans2025stable, kumar2024high} learn to compact audio latent and then to reconstruct.
% Both those vocoders and VAEs are purely based on acoustic representations, which are oriented toward reconstruction; in contrast, the SemanticVocoder utilizes semantic latents and is oriented toward generation.

\textbf{Text to audio generation}
TTA models accept text descriptions and generate corresponding audio.
Previous studies, such as AudioLDM2~\cite{liu2023audioldm2}, TangoFlux~\cite{hung2024tangoflux}, and Make-An-Audio~\cite{huang2023make1}, adopt a two-stage paradigm combining a VAE with a latent diffusion model.
The diffusion model generates text-conditioned latents within the VAE latent space, which are subsequently decoded into waveforms by the VAE decoder. 
This framework has become the mainstream approach for text-to-audio generation in recent years, and our work represents one of the pioneering attempts to build an audio generation framework that dispenses with VAEs entirely.

\textbf{Semantic audio encoder}
Semantic audio encoders learn high-level semantic representations from large-scale audio datasets, prioritizing semantic content over low-level acoustic details.
These latent representations demonstrate strong performance on various audio understanding tasks, such as classification, event detection, and audio-text retrieval, by capturing intrinsic semantic structure.
Typical semantic audio encoders include:
(1) Contrastive Language-Audio Pre-training (CLAP) models~\cite{laionclap2023, elizalde2024natural_msclao2023}, which adopt contrastive learning to align paired audio and text representations while distancing mismatched pairs;
(2) supervised pre-trained encoders~\cite{kong2020panns}, which are trained on large annotated datasets with full supervision to optimize classification objectives and learn task-oriented semantics;
(3) Masked Autoencoder (MAE) based encoders~\cite{dinkel2024scaling_dasheng}, which adopt a self-supervised masked reconstruction paradigm. 
By randomly masking and recovering parts of audio spectrograms or latent features, they capture structured semantic patterns without manual annotations.
We adopt the MAE semantic encoder in our work, as its reconstruction objective is more closely aligned with our waveform generation task compared to the other two.

\section{Methodology: generative semantic vocoder}
\begin{figure}[htbp]
\centerline{\includegraphics[width=\linewidth]{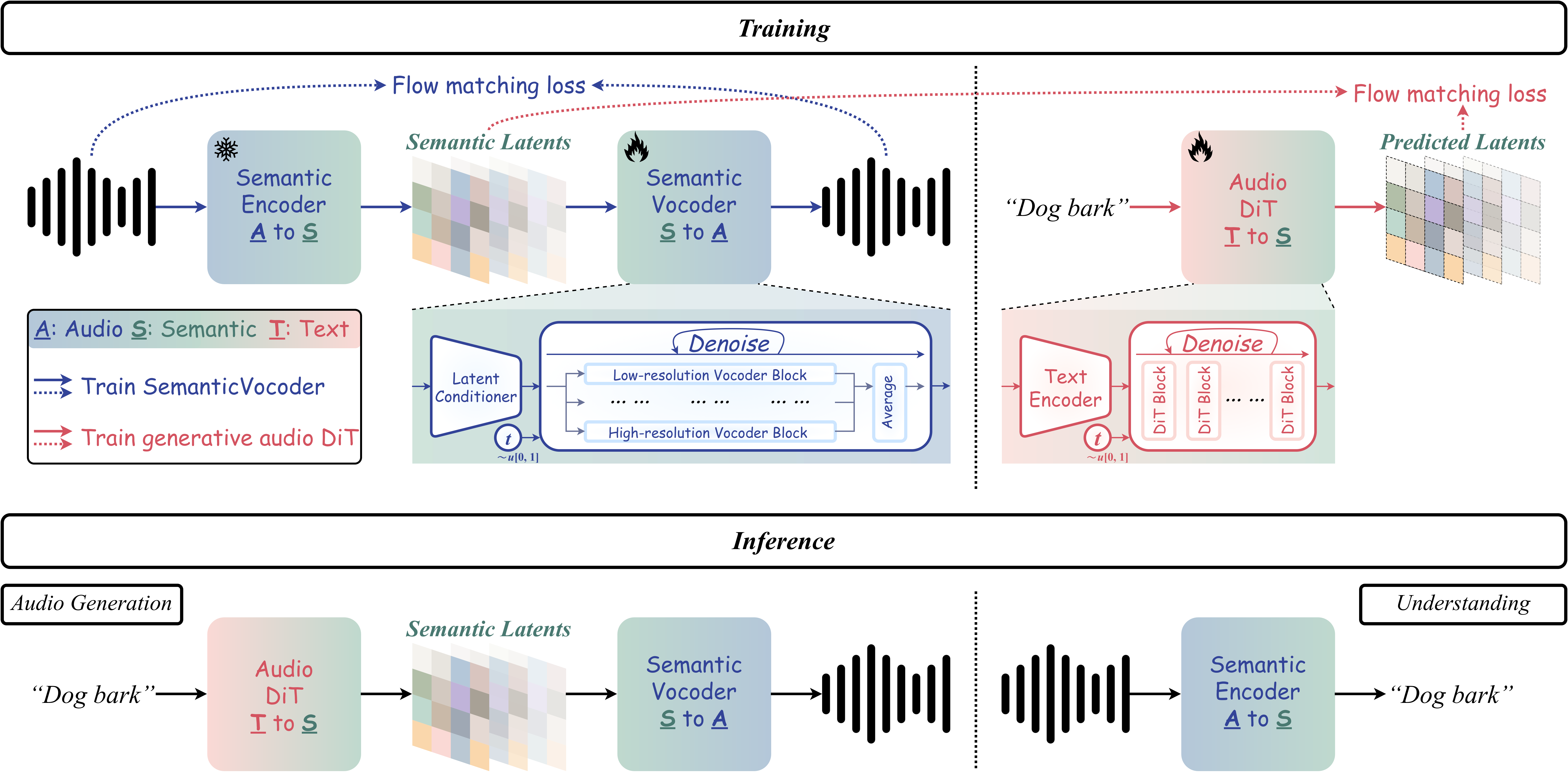}}
\caption{
An overview of SemanticVocoder training, downstream TTA training, and downstream task inference.
{\color[HTML]{324398}\bf{($\rightarrow$Blue arrow)}}
SemanticVocoder training: the input audio is fed into a semantic encoder to extract semantic latents, which serve as conditions to train the flow-matching network for waveform prediction.
{\color[HTML]{D45360}\bf{($\rightarrow$Red arrow)}}
Generative audio DiT training: the input text is processed by a text encoder to obtain textual features, which are used to train the DiT model for generating semantic latents.
{\bf{($\rightarrow$Black arrow)}}
Downstream task inference: equipped with SemanticVocoder, both audio generation and understanding tasks can be performed within the same semantic latent space.
}
\label{fig:pipe}
\end{figure}

SemanticVocoder innovatively generates waveforms directly from semantic latents.
As illustrated in the upper-lef part of Figure~\ref{fig:pipe}, it consists of:  
(1) a flow-matching based training strategy;
(2) a semantic encoder that extracts semantic latents from input audio;
and (3) a generative backbone for waveform prediction.

\subsection{Flow estimator}
\label{sec:flow_matching}
% Conventional VAEs typically rely on feed-forward decoder for waveform reconstruction, which presents difficulties when operating on semantic latents, since semantic latents mainly retain high-level semantic information while discarding fine-grained acoustic details.
Conventional VAEs are not straightforward for reconstruction from semantic latents, as these latents primarily preserve high-level semantic information while discarding fine-grained acoustic details, thus leading to reconstruction distortions.
To address this issue, we propose a generative approach to synthesize waveforms from semantic latents via a flow estimator.
Vanilla flow matching learns a velocity field $v_t(x_t)$ that transforms a noise distribution $p_0(x) = \mathcal{N}(0, \sigma^2 I)$ to the target distribution $p_1(x)$ over a continuous time interval $t \in [0, 1]$. 
The trajectory of a sample $x_t$ follows:
\begin{equation}
\frac{dx_t}{dt} = v_t(x_t), \quad x_0 \sim p_0(x), \quad x_1 \sim p_1(x)
\end{equation}
where $x_t = (1-t)x_0 + tx_1$ denotes the linear interpolation between the initial noise $x_0$ and target data $x_1$.
The training objective is to minimize the Mean Squared Error loss between the predicted velocity field $\hat{v}_t(x_t; \theta)$ and the ground-truth velocity $v_t^* = x_1 - x_0$:
\begin{equation}
\mathcal{L}_{\text{FM-velocity}} = \mathbb{E}_{t \sim \mathcal{U}(0,1), x_0 \sim p_0, x_1 \sim p_1} \left[ \left\| \hat{v}_t(x_t; \theta) - v_t^* \right\|_2^2 \right]
\end{equation}
We adopt the improved strategies introduced in Flow2GAN~\cite{yao2025flow2gan} for better waveform generation.
An energy-aware loss scaling method is used to prioritize perceptually critical segments.
Furthermore, the model $\hat{x}_1(x_t; \theta)$ directly predicts the target clean data $x_1$, since vanilla flow matching with velocity estimation tends to suffer from instability and noise amplification in low-energy audio regions:
%silent or 
\begin{equation}
\mathcal{L}_{\text{FM-data}} = \mathbb{E}_{t \sim \mathcal{U}(0,1), x_0 \sim p_0, x_1 \sim p_1} \left[ \left\| \hat{x}_1(x_t; \theta) - x_1 \right\|_2^2 \right]
\end{equation}
\subsection{Semantic encoder}
We employ a pretrained MAE encoder~\cite{dinkel2024scaling_dasheng} to extract semantic latents, as its masked predictive reconstruction paradigm is conceptually analogous to the waveform generation paradigm of SemanticVocoder.
During training, the MAE first converts the audio signal into a spectral representation, splits it into patches, and randomly masks a large fraction of these patches.
Only the unmasked patches are fed into the encoder to capture global structural patterns and long-range contextual dependencies. 
A lightweight decoder then reconstructs the masked patches. 
This training mechanism enables the encoder to learn generalizable and intrinsic semantic patterns that go beyond low-level acoustic details.
In practice, the decoder is discarded, and the encoder is used to encode the input audio without masking, yielding implicit features rich in high-level semantic information that serve as semantic latents $L_{\text{semantic}}$:
\begin{equation}
L_{\text{semantic}} = \mathcal{E}_{\text{MAE}}( \text{Patchify}(\text{Mel}(x)))
\end{equation}
\subsection{Generative backbone}
The generative backbone follows the design of traditional acoustic vocoders~\cite{siuzdak2023vocos, yao2025flow2gan, liu2024rfwave} and outputs waveforms conditioned on audio latents.
It consists of a latent conditioner and a waveform generator.

\subsubsection{Latent conditioner}
\label{sec:latents_encoder}
To better integrate the information from semantic latents, we employ a latent conditioner to further encode and map the semantic latents $L_{\text{semantic}}\in\mathbb{R}^{B\times D\times T}$ into a suitable feature space, resulting in $L'_{\text{semantic}} \in \mathbb{R}^{B\times D'\times T}$.
%Periodwave,
Specifically, the latent conditioner begins with a 1D convolutional layer to project features to the target channel size, followed by BiasNorm normalization and a stack of ConvNeXt~\cite{liu2022convnet} blocks.
%A ConvNeXt~\cite{liu2022convnet} module is utilized, which begins with a 1D convolutional layer to project features to the target channel size, followed by BiasNorm normalization and a stack of ConvNeXt blocks.
Each ConvNeXt block consists of a grouped convolution for spatial mixing, two pointwise convolutions for channel mixing with PReLU activation in between, and a residual connection for gradient flow.
\begin{equation}
L'_{\text{semantic}} = \mathcal{E}_{\text{conditioner}}(L_{\text{semantic}})
\end{equation}

\subsubsection{Waveform generator}
Waveform reconstruction from complex spectrograms can be effectively achieved via the Inverse Short-Time Fourier Transform (iSTFT)~\cite{siuzdak2023vocos, liu2024rfwave}.
Therefore, we take the Short-Time Fourier Transform (STFT) coefficients $x_{\text{coef}}$ as the network prediction target, which is further converted into the target waveform $x_1$ through iSTFT.
Specifically, a flow matching model $\mathcal{FM}_\text{SemanticVocoder}$ is employed to predict $x_{\text{coef}}$ based on the time interval $t \in \mathcal{U}[0, 1]$, the noisy data $x_t$, and the conditional semantic latent features $L'_{\text{semantic}}$:
\begin{equation}
x_{\text{coef}} = \mathcal{FM}_{\text{SemanticVocoder}}\left(x_t, t, L'_{\text{semantic}}\right)
\end{equation}
To enable $\mathcal{FM}$ to adapt to audio of varying complexity, we employ $R$ parallel ConvNeXt~\cite{liu2022convnet} branches to predict STFT coefficients at different spectral resolutions. 
The final waveform is obtained by averaging their outputs:
\begin{equation}
x_1 = \frac{1}{R} \sum_{r=1}^{R} \text{iSTFT}_R(x_{\text{coef-}R\text{th}})
\end{equation}
Each stacked ConvNeXt block shares a similar structure to that in Section \ref{sec:latents_encoder}, but is adapted for flow-matching timestep and conditional information, thus modulating the intermediate variable $x_{\text{hidden}}$.
In detail, the timestep information $t$ is embedded as $t_{\text{emb}}$ using sinusoidal positional encoding and processed through a MLP (Multilayer Perceptron, denotes as $\mathcal{P}$).
The embedded timestep features are multiplied, while the conditional features $L'_{\text{semantic}}$ are subsequently added.
\begin{equation}
t_{\text{emb}} = \mathcal{P}_{\text{time}}\left(\text{SinPE}(t)\right)
\label{eqt:time}
\end{equation}
\begin{equation}
x_{\text{hidden}} = x_{\text{hidden}}  \odot (1 + t_{\text{emb}})  +  \mathcal{P}_{\text{latents}}\left(L'_{\text{semantic}}\right) 
\end{equation}

\section{Downstream tasks}
SemanticVocoder establishes a direct mapping from semantic latents to waveforms, facilitating the integration of these latents into downstream text-to-audio generation.
%Furthermore, by comparing semantic latents with traditional VAE acoustic latents in downstream understanding tasks, 
Furthermore, through comparison in downstream understanding tasks, we demonstrate that semantic latents exhibit stronger discriminability and clearer semantic structure, which benefits audio generative modeling.

\subsection{Text to audio generation}
We employ a Diffusion Transformer (DiT)~\cite{peebles2023scalable_dit} for TTA to verify the benefits of introducing semantic latents, as shown in Figure~\ref{fig:pipe}.
It uses the vanilla flow-matching strategy described in Section~\ref{sec:flow_matching}.
Its prediction target is changed from traditional VAE latents to semantic latents $L_{\text{semantic}}$,  which are then converted into waveforms by SemanticVocoder.
Specifically, a text encoder is utilized to extract textual embeddings $c_{\text{emb}}$ from caption $c$.
The DiT backbone employs a stack of DiT blocks, which take text embeddings $c_{\text{emb}}$, timestep embeddings $t_{\text{emb}}$ (Equation~\ref{eqt:time}), and noisy latents $L_{\text{noisy}}$ as inputs.
\begin{equation}
L_{\text{semantic}}=\mathcal{FM}_{\text{audioDiT}}\left(L_{\text{noisy}}, t, c\right)
\end{equation}
%: 
% \begin{equation}
% c_{\text{emb}} = \mathcal{E}_{\text{text}}(c)
% \end{equation}
Each block utilizes AdaLN (adaptive layer normalization) to incorporate the timestep information and adopts cross-attention to fuse $c_{\text{emb}}$, thereby modulating the intermediate predicted latent $L_{\text{hidden}}$:
\begin{equation}
 L_{\text{hidden}} = \text{AdaLN}\left(t_{\text{emb}},   L_{\text{hidden}} \right)
\end{equation}
\begin{equation}
  L_{\text{hidden}} =\text{CrossAttn}\left(c_{\text{emb}},   L_{\text{hidden}}   \right) + L_{\text{hidden}}
\end{equation}

%The DiT uses the vanilla flow-matching strategy described in Section~\ref{sec:flow_matching}.
%The predicted latent output is further mapped to the waveform domain via SemanticVocoder.

\begin{figure}[thbp]
\centerline{\includegraphics[width=\linewidth]{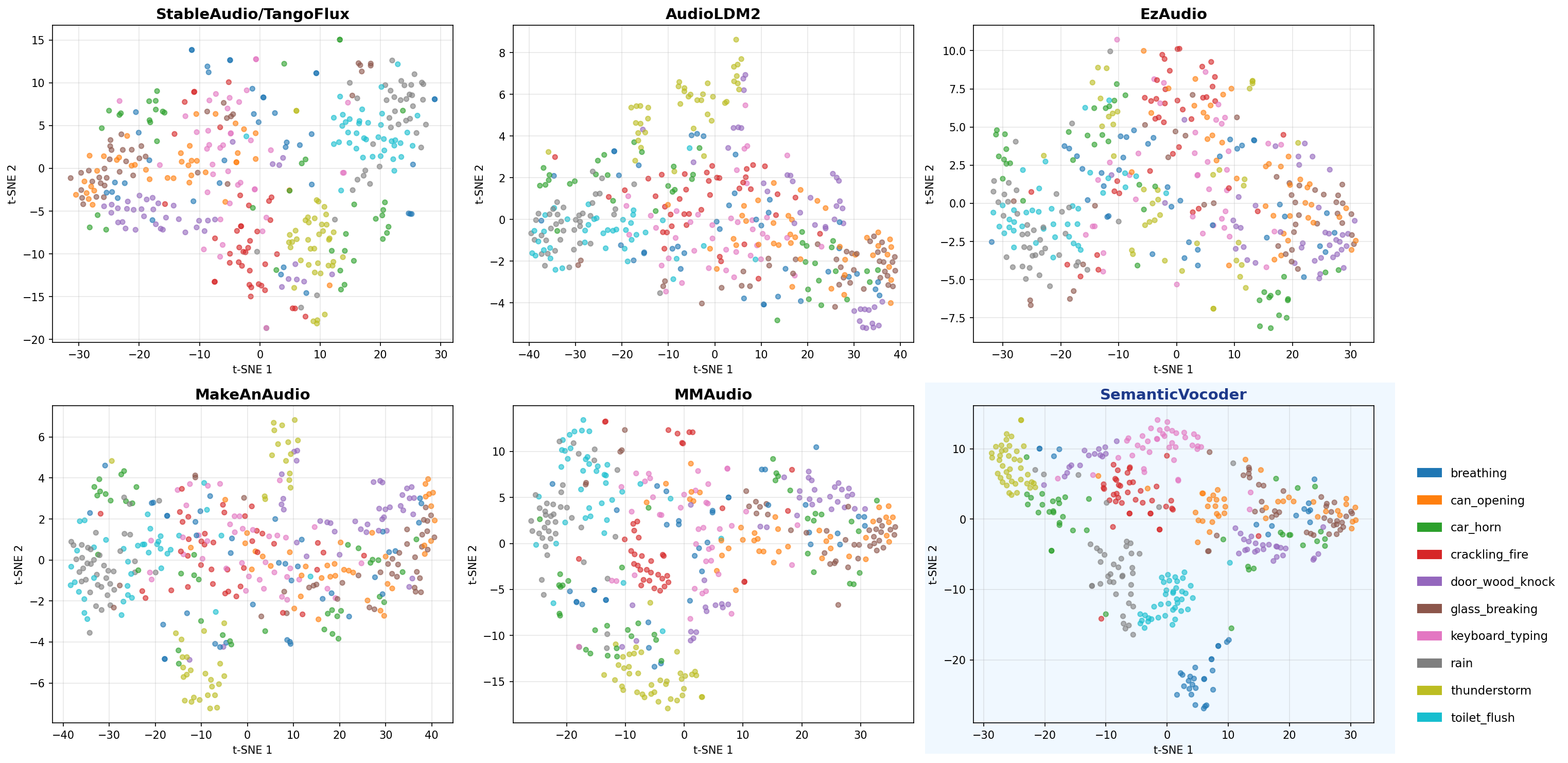}}
\caption{
%Visualization of different latents on HEAR-esc50, where the 10 most frequent categories are presented.
%Semantic latents exhibit stronger discriminative structure and semantic disentanglement compared to baseline models.
Visualization of different latents on HEAR-ESC50, where the 10 most frequent categories are presented.
Each audio feature is aggregated by mean pooling along the temporal axis and projected into 2D space via t-SNE.
Compared to VAE acoustic latents used in baseline models, semantic latents exhibit a more discriminative structure and superior semantic disentanglement.
}
\label{fig:emb}
\end{figure}

\subsection{Audio understanding}
To demonstrate why semantic latents benefit audio generation, we compare their performance with the original VAE acoustic latents on downstream understanding tasks.
We adopt the HEAR benchmark~\cite{turian2022hear} for evaluation, which trains separate MLPs on fixed latents for various understanding tasks, including multi-class classification, multi-label classification, and event detection.
%We evaluate the semantic discriminability of the latents on multi-class classification, multi-label classification, and event detection tasks.

\section{Implementation}

\subsection{Datasets}
\textbf{AudioSet}~\cite{gemmeke2017audio_audioset} A large-scale human-labeled audio event dataset containing about 2 million 10-second clips from YouTube videos, with annotations across 527 hierarchical audio event categories. 
%The dataset covers diverse sound types, including human vocalizations, animal sounds, environmental noises, musical instruments, and mechanical sounds. 
\textbf{AudioCaps}~\cite{kim2019audiocaps} A pioneering audio captioning dataset built upon AudioSet, which consists of about 50K audio clips (10 seconds each) paired with human-written natural language captions. 
\textbf{WavCaps}~\cite{mei2024wavcaps} As large-scale weakly-labeled audio captioning dataset. 
It comprises approximately 400K audio-caption pairs collected from 4 sources.
%: AudioSet (strongly labeled), BBC Sound Effects, FreeSound, and SoundBible. 
Raw web-harvested descriptions are refined via a three-stage pipeline to filter noise and generate human-like captions. %ChatGPT-assisted 

\subsection{Models setup}
\textbf{SemanticVocoder}
We employ the MAE-based $dasheng\_base$~\cite{dinkel2024scaling_dasheng} as the semantic encoder due to its strong downstream performance.
%It operates at a sampling rate of 16 kHz and outputs 768-dimensional features at 25 frames per second.
The latent conditioner employs a 4-layer ConvNeXt with a hidden dimension of 512.
We employ three parallel 8-layer ConvNeXt branchs as the flow-matching backbone at different resolutions, corresponding to STFT hop lengths $(320, 160, 80)$ and ConvNeXt hidden dimensions $(768, 512, 384)$, respectively.
They target the STFT coefficients of 24 kHz audio, and the final output waveform is obtained by averaging their outputs.
SemanticVocoder is trained on AudioSet for 270 epochs with the ScaledAdam optimizer~\cite{yaozipformer_scaledOpt}, with each audio clipped into 1.6-second segments. 
We use a batch size of 1440. 
The learning rate starts at $3.5\times 10^{-3}$, linearly warmed up to $3.5\times 10^{-2}$ over the first 500 steps, and decays using the Eden2 scheduler after 7500 steps.
During inference, an Euler ODE solver is used with a step size of 200.

\textbf{Text to audio generation}
The $Flan\_T5\_large$~\cite{chung2024scaling_flant5} is employed as the text encoder to extract text embeddings.
We utilize 24-layer canonical DiT blocks with a hidden dimension of 1024 and the AdaLN-SOLA fusion mechanism~\cite{xu2025uniflow, hai2024ezaudio}.
%We utilize the canonical DiT block with the AdaLN-SOLA fusion mechanism~\cite{xu2025uniflow, hai2024ezaudio}.
%Specifically, the base and large versions adopt a hidden dimension of 1024 and 1536, respectively, with 24 and 28 transformer blocks.
%The DiT is pre-trained on the union of WavCaps and AudioCaps for 40 epochs, followed by fine-tuning on AudioCaps for 300 epochs using a batch size of 32
The DiT is pretrained on the combined WavCaps and AudioCaps datasets for 40 epochs, followed by fine-tuning on AudioCaps for 300 epochs, using a batch size of 32, the AdamW optimizer, and a learning rate of $5\times 10^{-5}$.
During inference, the Euler ODE solver with 100 steps is used, and the classifier-free guidance scale is set to $3.5$.

\textbf{Audio understanding}
We follow the standard protocol of the HEAR benchmark~\cite{turian2022hear}.
Three highly relevant audio understanding tasks are selected:
HEAR-DCASE2016 (multi-label detection), HEAR-ESC50 (multi-class classification), and HEAR-FSD50k (multi-label classification).
SemanticVocoder uses semantic latents, while other systems employ acoustic latents sampled from VAE encoders.
%All evaluation settings follow the HEAR pipeline.

% \begin{table}[t]
% \centering
% \resizebox{0.8\linewidth}{!}{  
% \begin{tabular}{lccccc}
% \toprule
% Model & FAD$_{\text{VGG}}$$\downarrow$ & FAD$_{\text{PANNs}}$ (FD)$\downarrow$ & KL$\downarrow$ & IS$\uparrow$ & CLAP$\uparrow$ \\
% \midrule
% GT          & -          & 0.524          & -          & -          & -          \\
% ezaudio     & 2.261670106 & 15.73771       & 1.32187    & 11.01828585 & 0.458      \\
% audioldm2   & 2.95744    & 24.38693       & 1.54529    & 9.104856569 & 0.395      \\
% tangoflux   & 2.33809    & 19.31881       & 1.18341    & 12.36549426 & 0.482      \\
% makeanaudio & 2.45508    & 18.16876       & 1.50552    & 8.81973528  & 0.406      \\
% affusion$\times$ & 2.21869  & 20.41009       & 1.2966     & 10.99224408 & 0.443      \\
% stable-audio & 3.40918   & 44.80568       & 2.15206    & 9.44628602  & 0.216      \\
% mmaudio     & 6.18442    & 18.63059       & 1.33715    & 12.14426304 & 0.416      \\
% \bottomrule
% \end{tabular}
% }
% \caption{Quantitative evaluation of audio generation models on core metrics. Lower values are better for FAD$_{\text{VGG}}$, FAD$_{\text{PANNs}}$, and KL; higher values indicate superior performance for IS and CLAP.}
% \label{tab:audio_gen_metrics}
% \end{table}

\subsection{Metrics}
For text-to-audio generation, we adopt common objective metrics~\footnote{AudioLDM Eval: \url{https://github.com/haoheliu/audioldm_eval}}, including FAD, FD, KL divergence, Inception Score (IS), and CLAP similarity~\footnote{ 
LAION-CLAP: \url{https://github.com/LAION-AI/CLAP} \\
   \phantom{LAI}MS-CLAP: \url{https://github.com/microsoft/CLAP}}.
Although SemanticVocoder is designed for generative tasks, its reconstruction performance remains comparable to reconstruction models.
We therefore also report reconstruction metrics~\footnote{DAC: \url{https://github.com/descriptinc/descript-audio-codec}}, including ViSQOL and Mel/STFT/Waveform loss.

%\subsection{Comparison}
% All evaluations are conducted on the AudioCaps test set using the first corresponding caption, ensuring fair comparison.
% Well-performing systems (EzAudio~\cite{hai2024ezaudio}, AudioLDM2~\cite{liu2023audioldm2}, TangoFlux~\cite{hung2024tangoflux}, MakeAnAudio~\cite{huang2023make1}, StableAudio~\cite{evans2025stable}, MMAudio~\cite{cheng2025mmaudio}) serve as comparative baselines. 

%We further conduct ablation studies thatreplacing the semantic latents in our framework with VAE latents or mel-spectrograms, to verify the superiority of semantic latents over conventional acoustic latents.
%or adopting a reconstruction-oriented training scheme~\cite{evans2025stable} for SemanticVocoder, to validate the necessity of the generative paradigm.

\section{Results}
This section presents the experimental results on text-to-audio generation, semantic representation capability, and reconstruction ability.
Additional details and ablation studies are provided in  \hyperref[sec:ablation]{Appendix}.
High-performing systems, including EzAudio~\cite{hai2024ezaudio}, AudioLDM2~\cite{liu2023audioldm2}, TangoFlux~\cite{hung2024tangoflux}, MakeAnAudio~\cite{huang2023make1}, StableAudio~\cite{evans2025stable}, and MMAudio~\cite{cheng2025mmaudio}, serve as comparative baselines. 

\begin{table}[t]
\centering
\resizebox{\linewidth}{!}
{  
\begin{tabular}{lcccccc}
\toprule
System & FAD$_{\text{VGG}}$ $\downarrow$ & FD$_{\text{PANNs}}$ $\downarrow$ & KL $\downarrow$ & IS $\uparrow$ & CLAP$_{\text{LAION}}$ $\uparrow$  & CLAP$_{\text{MS}}$ $\uparrow$\\
\midrule
EzAudio~\cite{hai2024ezaudio}     & 2.262 & 15.738       & 1.322    & 11.018 & 0.458   &0.423   \\
AudioLDM2~\cite{liu2023audioldm2}    & 2.957    & 24.387       & 1.545    & 9.105 & 0.395  &0.556     \\
TangoFlux~\cite{hung2024tangoflux}    & 2.338    & 19.319       & 1.183    & \textbf{12.365} & \textbf{0.482}  &0.468    \\
MakeAnAudio~\cite{huang2023make1}  & 2.455    & 18.169       & 1.506    & 8.820  & 0.406  & \textbf{0.574}    \\
%Affusion~\cite{} $\times$ & 2.219  & 20.410       & 1.297     & 10.992 & 0.443  &0.617    \\
StableAudio~\cite{evans2025stable} & 3.409   & 44.806       & 2.152    & 9.446  & 0.216   &0.345   \\
MMAudio~\cite{cheng2025mmaudio}     & 6.184    & 18.631       & 1.337    & 12.144 & 0.416   &0.446   \\
\midrule
SemanticVocoder &\textbf{1.709	} &\textbf{12.823	}	&\textbf{1.154	}	&11.286		&0.454 &0.557\\
%SemanticVocoder &\textbf{1.820} &\textbf{13.114}	&\textbf{1.148}	&11.386	&0.455 &0.556\\
%SemanticVocoder &\textbf{1.955} &	\textbf{13.175}	& \textbf{1.155}	&11.403 &	0.453 &  0.555\\
%SemanticVocoder     & \textbf{1.836} &\textbf{12.297}&1.238&	11.727 &	0.445 &0.554\\
%SemanticVocoder &1.987&	11.973&	1.209&	10.996&	0.451& 0.557 \\
%SemanticVocoder &\textbf{1.829}	&\textbf{12.964}	&1.249&	11.928&	0.448 &0.554\\
% \quad with VAE latents &2.983 &	15.768	 &1.446	&9.976	  &0.421 &0.378 \\  
% \quad with Mel latents &5.896	 &24.745	 &2.045	 &7.171	&0.379  & 0.404\\ 

\bottomrule
\end{tabular}
}
\caption{Text-to-audio generation performance on AudioCaps.
The best FAD and FD scores reflect that the audio distribution generated by SemanticVocoder is closest to that of the reference audio.
% ``with'' denotes that the semantic latents are replaced with VAE latents or mel-spectrograms in ablation studies.
}
\label{tab:audio_gen_metrics}
\end{table}

\subsection{Text to audio generation performance}
\label{sec:tta_result}
The quantitative evaluation results for text-to-audio generation are presented in Table \ref{tab:audio_gen_metrics}.
All evaluations are conducted on the AudioCaps test set using the first corresponding caption, ensuring fair comparison.
Our SemanticVocoder achieves the lowest FAD$_{\text{VGG}}$ and FD$_{\text{PANNs}}$ scores, outperforming all baseline models while maintaining competitive performance on other metrics.
As FAD and FD measure the distribution distance between generated and reference audio, it indicates that SemanticVocoder produces audio closer to the real distribution than VAE-based systems.
% The ablation studies also show that replacing semantic latents with conventional acoustic latents (VAE or Mel for acoustic vocoder) causes performance degradation.

To summarize the underlying reasons, compared with directly predicting low-level acoustic latents from text, high-dimensional semantic latents serve as more suitable targets for generative modeling.
The VAE acoustic latents contain excessive details, rendering second-stage text-to-latent modeling difficult to optimize.
In other words, in conventional VAE-based LDM frameworks, text-to-latent prediction is challenging, whereas latent-to-waveform generation remains relatively straightforward.
\textbf{In contrast, SemanticVocoder balances optimization difficulty across these two stages}.
These results demonstrate the effectiveness of substituting traditional acoustic latents with semantic latents.

\begin{table*}[t]
\centering
\resizebox{\textwidth}{!}{
\begin{tabular}{l c c c c c}
\toprule
System & Dimension & DCASE2016 $\uparrow$&  ESC50 $\uparrow$& FSD50k $\uparrow$\\
\midrule

StableAudio~\cite{evans2025stable}/TangoFlux~\cite{hung2024tangoflux} & 64 & 60.523 & 42.450& 21.134 \\
EzAudio~\cite{hai2024ezaudio} & 128 & 21.616 & 22.850 & 10.768 \\
AudioLDM2~\cite{liu2023audioldm2} & 8*16 & 20.821 & 36.150 & 16.221 \\
%affusion & 4*32 & & 61.731 & 37.800 & 19.511 \\
MakeAnAudio~\cite{huang2023make1}  & 4*10& 68.771 & 35.250 & 15.694 \\
MMAudio~\cite{cheng2025mmaudio} & 40 & 56.770 & 35.600 & 15.831 \\
\midrule
SemanticVocoder & 768 & \textbf{93.690}
 & \textbf{83.400} & \textbf{51.583} \\
\bottomrule
\end{tabular}
}
\caption{
Evaluation scores of acoustic and semantic latents on the HEAR audio understanding benchmark.
TangoFlux adopts the same VAE as StableAudio.}
\label{tab:understanding}
\end{table*}

\subsection{Audio understanding capability}
The semantic discriminability performance on the HEAR benchmark is shown in Table~\ref{tab:understanding}.
Benefiting from the powerful pre-trained semantic encoder~\cite{dinkel2024scaling_dasheng}, the semantic latents significantly outperform the acoustic latents on all three understanding tasks.
As also observed from Figure~\ref{fig:emb}, semantic latents exhibit a clearer distribution across different sound events.
This reveals the drawback of original VAE latents, which prioritize compression and reconstruction, yielding low-level acoustic encodings that lack structured semantic discriminability.
Furthermore, these results illustrate that SemanticVocoder opens up an opportunity to unify audio generation and  understanding within a single latent space.

\begin{table*}[t]
\centering
%\resizebox{1.0\textwidth}{!}
{
\begin{tabular}{l c c c c c}
\toprule
System &  Sample Rate &ViSQOL $\uparrow$ & Mel $\downarrow$ & STFT $\downarrow$ & Waveform $\downarrow$ \\
\midrule
StableAudio~\cite{evans2025stable}/TangoFlux~\cite{hung2024tangoflux} & 44.1K & 4.381 & 0.766 & \textbf{2.001} & 0.047 \\
EzAudio~\cite{hai2024ezaudio} &24K& \textbf{4.550} & 1.115 & 2.267 & \textbf{0.038} \\
AudioLDM2~\cite{liu2023audioldm2} &16K &3.514 & 1.488 & 3.332 & 0.084 \\
MakeAnAudio~\cite{huang2023make1} &16K& 3.214 & 1.498 & 3.831 & 0.089 \\
%affusion & 3.289 & 1.613 & 3.672 & 0.090 \\
MMAudio~\cite{cheng2025mmaudio}&44.1K & 4.481 & \textbf{0.620} & 2.003 & 0.099 \\
\midrule
SemanticVocoder &24K& 3.239 & 1.304 & 3.283 & 0.096 \\
\bottomrule
\end{tabular}
}
\caption{Reconstruction evaluation on AudioCaps test set.
Despite SemanticVocoder being optimized for generation based on semantic information, it exhibits comparable reconstruction performance.
TangoFLux adopts the same VAE from StableAudio.
}
\label{tab:audio_reconstruction_metrics}
\end{table*}

\subsection{Audio reconstruction performance}
Although SemanticVocoder is primarily designed for generative tasks, we additionally evaluate its reconstruction performance on the AudioCaps test set, as reported in Table \ref{tab:audio_reconstruction_metrics}.
SemanticVocoder is comparable to dedicated reconstruction-focused baselines. 
%This demonstrates that semantic latents not only can benefit generation but also might support robust audio reconstruction.
This demonstrates that semantic latents can benefit not only generation but also robust audio reconstruction.
%These results collectively confirm that discarding VAE acoustic latents and adopting semantic encoder latents is a promising direction for audio generation, as it addresses the issue of entangled event semantics in VAE latents and enables synergistic improvement across generation, understanding, and reconstruction tasks.

\section{Discussion}
\textbf{Advantages}
SemanticVocoder breaks away from the dependence of audio generation on VAEs, thereby alleviating several inherent limitations.
It rebalances the optimization difficulty across the two-stage pipeline: text-to-latent and latent-to-waveform generation.
In traditional frameworks, modeling the text-to-acoustic latent mapping is notoriously challenging, whereas converting acoustic latents to waveforms remains relatively straightforward. 
In contrast, SemanticVocoder fundamentally rebalances this trade-off by enabling the generative model to focus on the simpler task of aligning text with high-level semantic latents, while shifting the burden of reconstructing fine-grained acoustic details to the vocoder.
Meanwhile, SemanticVocoder aligns the optimization objectives of both stages toward generation.
Finally, by employing semantic latents as an intermediate anchor, the two stages become mutually independent, and can thus be trained separately and are inherently plug-and-play.

% In contrast, SemanticVocoder fundamentally rebalances this by allowing the generative model to focus on the easier task of aligning text with high-level semantic latents, shifting the burden of reconstructing complex acoustic details to the vocoder.
\textbf{Limitations}
SemanticVocoder still has several limitations.
Its performance depends on the capability of the pretrained semantic encoder to abstract semantic structures.
It is constrained by audio length: the model currently cannot generate long-form audio.
The objective metrics such as FAD and FD cannot reflect all aspects of audio generation quality, necessitating subjective evaluation to further assess generation quality from a human perception perspective.
%; thus, subjective experiments are still needed to further verify the quality of generated audio from a human perception perspective.

\section{Conclusion}
In this paper, we propose SemanticVocoder, a novel generative vocoder that directly synthesizes audio waveforms from semantic latents, departing from conventional VAE-based acoustic latents.
By discarding the conventional VAE module that focuses on low-level acoustic compression and reconstruction, our framework effectively mitigates issues of semantic entanglement and weak discriminability inherent in traditional latent spaces.
We adopt a generative paradigm to mitigate the waveform distortion caused by feed-forward reconstruction networks when recovering signals from semantic latents.
Downstream experiments on text-to-audio generation demonstrate that SemanticVocoder outperforms baseline systems, as evidenced by the lowest FAD and FD scores, which indicate closest alignment between generated audio and the real distribution.
Meanwhile, evaluations on semantic understanding benchmarks reveal that the introduced semantic latents exhibit a highly discriminative structure and clear semantic separation, demonstrating the superiority of semantic latents over acoustic latents.
Although primarily optimized for generation, SemanticVocoder maintains competitive reconstruction capability.
This underscores the flexibility and generality of semantic representations enabled by SemanticVocoder.
Furthermore, it balances optimization difficulty across two-stage models and exhibits inherent plug-and-play properties.
Overall, SemanticVocoder not only offers an effective alternative to VAE-centric audio generation pipelines but also presents a promising avenue toward unifying audio generation and understanding within a shared semantic space.

\bibliographystyle{IEEEtran}
\bibliography{ref}

\newpage
\appendix

\section{Ablation studies}
%%%%%%%%%%%%%%%%%%%%%%%%%%%%%%%%%%%%%%%%%%%%%%%%%%%%%%%%%%%%
\label{sec:ablation}
We conduct ablation studies comparing semantic latents with conventional acoustic latents within the same training framework.
We employ two acoustic latent representations: VAE latents and Mel latents.
For VAE latents, we adopt the pre-trained VAE from EzAudio~\cite{hai2024ezaudio} and perform text-to-latent generation within this latent space.
For Mel latents, the model predicts the Mel spectrograms, which are subsequently converted to waveforms via the BigVGAN~\cite{lee2022bigvgan} acoustic vocoder.
All text-to-latent models are trained exclusively on AudioCaps.
As shown in Table~\ref{tab:ablation_latents}, these results are consistent with those in Section~\ref{sec:tta_result}, demonstrating that semantic latents achieve superior performance relative to acoustic latents.
This superiority arises because acoustic latents encode low-level signal properties, whereas semantic latents capture high-level structural information more conducive to generative modeling.

\begin{table}[htbp]
\centering
\resizebox{\linewidth}{!}
{  
\begin{tabular}{lcccccc}
\toprule
Latent Type & FAD$_{\text{VGG}}$ $\downarrow$ & FD$_{\text{PANNs}}$ $\downarrow$ & KL $\downarrow$ & IS $\uparrow$ & CLAP$_{\text{LAION}}$ $\uparrow$  & CLAP$_{\text{MS}}$ $\uparrow$\\
\midrule
VAE acoustic latents &2.449 &16.358	 &1.531	&10.250	  &0.419 &0.391 \\  
Mel acoustic latents &5.087 &22.295	 &1.790	 &8.475	&0.410  &0.421 \\ 
Semantic latents &\textbf{1.956} &\textbf{14.110}	&\textbf{1.292}	&\textbf{11.038}	&\textbf{0.438} &\textbf{0.541}\\
\bottomrule
\end{tabular}
}				
\caption{Ablation studies on replacing semantic latents with VAE latents or mel-spectrograms.
}
\label{tab:ablation_latents}
\end{table}
				
% \begin{table}[htbp]
% \centering
% \resizebox{\linewidth}{!}
% {  
% \begin{tabular}{lcccccc}
% \toprule
% Latent Type & FAD$_{\text{VGG}}$ $\downarrow$ & FD$_{\text{PANNs}}$ $\downarrow$ & KL $\downarrow$ & IS $\uparrow$ & CLAP$_{\text{LAION}}$ $\uparrow$  & CLAP$_{\text{MS}}$ $\uparrow$\\
% \midrule

% Semantic &\textbf{1.820} &\textbf{13.114}	&\textbf{1.148}	&11.386	&0.455 &0.556\\
% VAE &2.983 &	15.768	 &1.446	&9.976	  &0.421 &0.378 \\  
% Mel &5.896	 &24.745	 &2.045	 &7.171	&0.379  & 0.404\\ 

% \bottomrule
% \end{tabular}
% }
% \caption{Quantitative evaluation of text to audio generation models on AudioCaps.
% ``with'' denotes that the semantic latents are replaced with VAE latents or mel-spectrograms in ablation studies.
% }
% \label{tab:ablation_latents}
% \end{table}

\section{Generative versus reconstructive approaches}

We further investigate alternative strategies to verify the necessity of a generative framework for latent-to-waveform generation.
For fair comparison, we employ the same text-to-latent model used in Table~\ref{tab:audio_gen_metrics}, since the introduction of semantic latents as an anchor renders our two-stage pipeline mutually independent and plug-and-play.
We adopt the training framework of the StableAudio~\cite{evans2025stable} VAE to reconstruct waveforms from semantic latents.
As shown in Table~\ref{tab:ablation_strategy}, employing a a reconstruction-based approach for latent-to-waveform conversion results in performance degradation, as traditional reconstruction methods suffer from audio distortion when dealing with semantic latents.

\begin{table}[htbp]
\centering
\resizebox{\linewidth}{!}
{  
\begin{tabular}{llccccc}
\toprule
Strategy & Architecture &FAD$_{\text{VGG}}$ $\downarrow$ & FD$_{\text{PANNs}}$ $\downarrow$ & KL $\downarrow$ & IS $\uparrow$ & CLAP$_{\text{LAION}}$ $\uparrow$  \\
\midrule

Reconstruction &VAE &4.781	 &18.962	 &1.212	 &8.240	 &0.403 \\
Generation&Flow-Matching   &\textbf{1.709} &\textbf{12.823}	&\textbf{1.154}	&\textbf{11.286}		&\textbf{0.454} \\

\bottomrule
\end{tabular}
}
\caption{
Text-to-audio generation results on AudioCaps.
The same text-to-semantic latent model is used, while different training strategies are adopted for the latent-to-waveform module.
}
\label{tab:ablation_strategy}
\end{table}
% \begin{table}[htbp]
% \centering
% \resizebox{\linewidth}{!}
% {  
% \begin{tabular}{llcccccc}
% \toprule
% Sstrategy & Architecture &FAD$_{\text{VGG}}$ $\downarrow$ & FD$_{\text{PANNs}}$ $\downarrow$ & KL $\downarrow$ & IS $\uparrow$ & CLAP$_{\text{LAION}}$ $\uparrow$  & CLAP$_{\text{MS}}$ $\uparrow$\\
% \midrule

% Reconstruction &VAE &4.781	 &18.962	 &1.212	 &8.240	 &0.403 &\textbf{0.564} \\
% Generation&Flow-Matching   &\textbf{1.709} &\textbf{12.823}	&\textbf{1.154}	&\textbf{11.286}		&\textbf{0.454} &0.557\\

% \bottomrule
% \end{tabular}
% }
% \caption{Quantitative evaluation of text to audio generation models on AudioCaps.
% ``with'' denotes that the semantic latents are replaced with VAE latents or mel-spectrograms in ablation studies.
% }
% \label{tab:ablation_feature}
% \end{table}

\section{Sampling steps and Class-Free guidance}
\begin{figure}[thbp]
\centerline{\includegraphics[width=\linewidth]{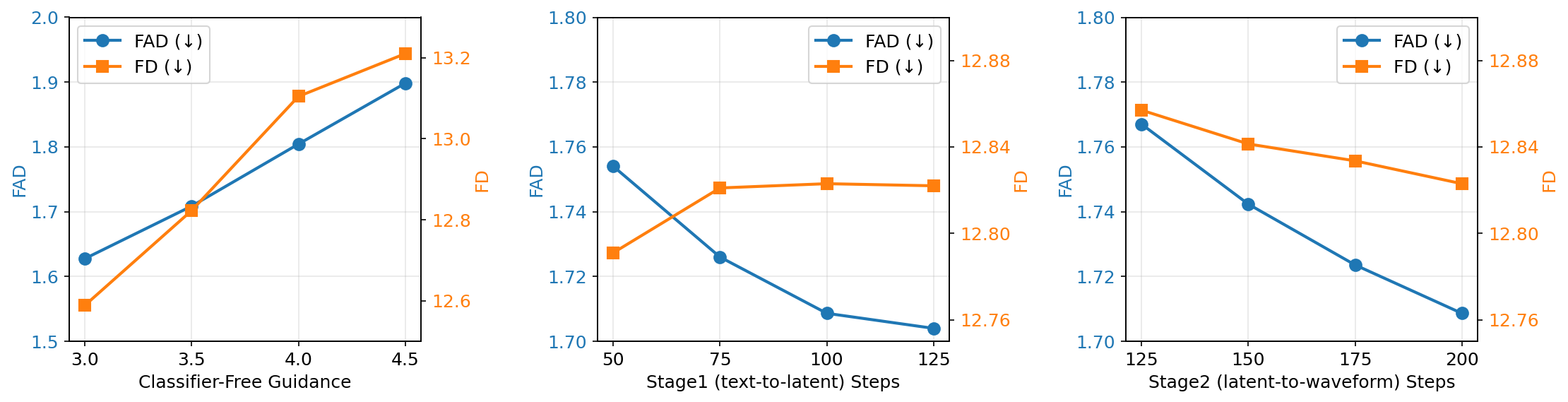}}
\caption{
The influence of inference steps and Class-Free guidance on TTA performance.
}
\label{fig:cfg_step}
\end{figure}

We investigate the effects of different numbers of inference steps and Class-Free Guidance (CFG) scale on the generation results, as shown in Figure~\ref{fig:cfg_step}.
It can be observed that CFG has a relatively large impact on model performance, while the performance varies only slightly across different sampling steps for both the text-to-latent and latent-to-waveform modules.

\section{Cross domain generation capability}
For cross-domain evaluation, we utilize the Clotho~\cite{drossos2020clotho} test set.
The first corresponding caption is used as the input prompt.
Each audio clip is truncated to 10 seconds.
It can be observed from Table~\ref{tab:clotho} that SemanticVocoder also achieves competitive performance and obtains the best FAD and FD scores.
\begin{table}[htbp]
\centering
\resizebox{\linewidth}{!}
{  
\begin{tabular}{lcccccc}
\toprule
System & FAD$_{\text{VGG}}$ $\downarrow$ & FD$_{\text{PANNs}}$ $\downarrow$ & KL $\downarrow$ & IS $\uparrow$ & CLAP$_{\text{LAION}}$ $\uparrow$  & CLAP$_{\text{MS}}$ $\uparrow$\\
\midrule
EzAudio~\cite{hai2024ezaudio}    &1.887	&20.499 &2.581 &10.476 &0.264 &0.473  \\
AudioLDM2~\cite{liu2023audioldm2} &2.677 &21.257 &\textbf{2.188}	&8.754 &0.289 &0.422  \\
TangoFlux~\cite{hung2024tangoflux} &2.503 &21.899 &2.462 &\textbf{10.833}	&0.318  &0.495  \\
MakeAnAudio~\cite{huang2023make1}  &3.368 &23.032 &2.631 &8.173 &0.236 &0.384  \\
StableAudio~\cite{evans2025stable}  &2.994 &40.731 &2.515	&9.168	 &0.285 &0.481  \\
MMAudio~\cite{cheng2025mmaudio}      &2.257	&20.143	&2.501 &9.223 &\textbf{0.356} &\textbf{0.515}  \\
\midrule
SemanticVocoder  &\textbf{1.833} &\textbf{18.690} &2.542 &9.651 &0.265 &0.465  \\

\bottomrule
\end{tabular}
}
\caption{Cross-domain evaluation results. 
Experiments are conducted on the Clotho test set, where the first caption is used as input and each audio clip is truncated to 10 seconds.
}
\label{tab:clotho}
\end{table}

\end{document}